\documentclass[10pt,amsmath,amssymb,aps,pre,letterpaper,twocolumn,floatfix,superscriptaddress]{revtex4-1}
\usepackage{graphicx,xcolor} 
\usepackage{bm}
\usepackage{textcomp}
\usepackage{gensymb}
\usepackage{array,multirow}
\newcolumntype{K}[1]{>{\centering\arraybackslash}m{#1}}
\usepackage{makecell}
\usepackage{tablefootnote}
\usepackage[english]{babel}
\usepackage{blindtext}
\usepackage{lipsum}
\usepackage{caption}
\usepackage{subcaption}
\usepackage{amssymb}
\usepackage{amsfonts}
\usepackage{amsmath}

\DeclareMathOperator{\csch}{csch}

\graphicspath{{./graph/}}
\renewcommand{\(}{\left(}
\renewcommand{\)}{\right)}


\begin{document}
	
	\title{Diffusive Exit Rates Through Pores in Membrane-Enclosed Structures}
	
	\author{Zitao Yang}
        \email{zitao.yang@ljcds.org}
	\affiliation{La Jolla Country Day School, La Jolla, California 92037}	
	
	\author{Elena F. Koslover}
	\email{ekoslover@ucsd.edu}
	\affiliation{Department of Physics, University of California, San Diego, San Diego, California 92093}
	\date{\today}
	\preprint{}

\begin{abstract}

The function of many membrane-enclosed intracellular structures relies on release of diffusing particles that exit through narrow pores or channels in the membrane. The rate of release varies with pore size, density, and length of the channel. We propose a simple approximate model, validated with stochastic simulations, for estimating the effective release rate from cylinders, and other simple-shaped domains, as a function of channel parameters. The results demonstrate that, for very small pores, a low density of channels scattered over the boundary is sufficient to achieve substantial rates of particle release. Furthermore, we show that increasing the length of passive channels will both reduce release rates and lead to a less steep dependence on channel density. Our results are compared to previously-measured local calcium release rates from tubules of the endoplasmic reticulum, providing an estimate of the relevant channel density responsible for the observed calcium efflux.
\end{abstract}


\maketitle
\newpage

\section{Introduction}
\label{sec:introduction}
The escape of diffusive particles from pores within membrane-enclosed structures occurs frequently within living cells. The particles can include ions, metabolites, proteins, and mRNAs. The enclosing membrane-bound domains may be cylindrical structures (such as the tubules of the peripheral endoplasmic reticulum (ER), neuronal axons, or fungal hyphae), spheres (as in the nuclear envelope or smaller vesicular organelles), or sheets (as in the perinuclear ER). Specific processes that rely on escape through narrow pores include Ca$^{2+}$ ion release from intracellular storage structures in the ER~\cite{basnayake2021nanoscale,okubo2020visualization,smith2009imaging,santulli2017intracellular} or lysosomes~\cite{yang2019release,patel2015evolution},  and transport of mRNA and proteins out of the nucleus via nuclear pores~\cite{bensidoun2021choosing,ma2013high}. 



The rate at which particles escape through the domain boundary is often described in terms of an effective membrane permeability~\cite{robertson1983lively,park2017maximizing}. 
For extensive membrane-enclosed structures, such as the peripheral ER network, the permeable region containing open channels may be highly localized. In this case, permeability within the release region can tune between regimes where long-range luminal transport versus local release rates are limiting for the overall leakage of particles out of the organelle~\cite{konno2021endoplasmic}.
 When the escape must occur through narrow pores or channels, the effective permeability depends on the pore size~\cite{park2017maximizing,schuss2007narrow} and density in the membrane. While the effect of membrane thickness and chemical properties on permeability have been thoroughly explored~\cite{phillips2012physical,robertson1983lively}, there are few general expressions for the dependence of permeability on the density of pores permitting escape.
 
  An analogous problem was considered in the classic paper by Berg and Purcell~\cite{berg1977physics}, focusing on the overall rate of diffusive particles being absorbed by an isolated sphere with small absorbing patches on its surface. This work demonstrated that only a small fraction of the sphere's surface area need be covered by pores to achieve near maximum diffusive flux. Similar results were also found for particles absorbed by patches on an isolated flat plane~\cite{berg2018random}. Here we seek to expand these results to compute the rate of escape of particles from pores and channels in enclosed domains of simple geometry (cylinders, spheres, and plates). Our primarily goal is to establish how the effective membrane permeability of cellular structures might vary with the density of the channels.

The leakage of particles through pores in a membrane-enclosed domain has also been considered as a generalization of a narrow escape problem, which seeks to compute the mean first passage time of an individual particle leaving through a small hole in the domain boundary~\cite{schuss2007narrow,benichou2008narrow,holcman2013control,cheviakov2012mathematical}. The inverse of this mean escape time, scaled by the number of particles in the domain, can be equated to the overall flux through the narrow pores~\cite{grigoriev2002kinetics,holcman2013control}.
Past work on the narrow-escape problem primarily focused on systems with one small absorbing window on the surface of the geometry.
First passage time expressions have been derived for circular disks and arbitrarily-shaped three dimensional domains~\cite{singer2006narrow1, singer2006narrow2, singer2006narrow3}.
A variety of more complex single-hole systems have been explored, such as diffusion through a funnel or bottleneck~\cite{holcman2011narrow,holcman2012brownian}, narrow escape through switching gates~\cite{ammari2011mean}, and narrow escape of interacting particles~\cite{agranov2018narrow}. In addition, asymptotic results are available for leakage through multiple small absorbing pores that are well-separated along the boundary~\cite{cheviakov2012mathematical}.

An additional complication in realistic biological structures is the need to pass through a channel of non-trivial length (greater than or comparable to the membrane thickness) after encountering a pore. Some work has been done on the contribution of bottle-neck length for narrow escape processes with a single outlet~\cite{holcman2012brownian}. However the diffusive flux of particles out of a domain with multiple pores that lead to channels of non-zero length has not been thoroughly explored.

In this work we expand the intuitively simple surface-sampling model presented in Berg's seminal work~\cite{berg1977physics,berg2018random} to derive the diffusive leakage rate as a function of pore density in several different domain geometries. We show that quite low pore densities can lead to very high flux, but that this steep dependence is substantially reduced when the pores lead to channels of appreciable length. Our aim is to provide simple expressions for the relationship between channel density, channel length, and effective membrane permeability, which can be leveraged to estimate expected fluxes of particles from biological structures.


\section{Results}
\label{sec:results}

\subsection{Low pore densities allow rapid leakage from cylinder}
\label{sec:model}


We begin by considering the diffusion of point-like particles in a cylinder with radius $R$. Such a cylinder can represent, for instance, a section of membrane-enclosed tubule in the peripheral ER ($R\approx 50$nm~\cite{schroeder2019dynamic}) or a portion of a neuronal axon ($R \approx 0.05-5\mu\text{m}$~\cite{perge2012axons}).
 The ends of the cylinder are treated as periodic boundary conditions, so that the cylinder represents a portion of a much longer tubule.

\begin{figure*}
 	\includegraphics[width=\textwidth]{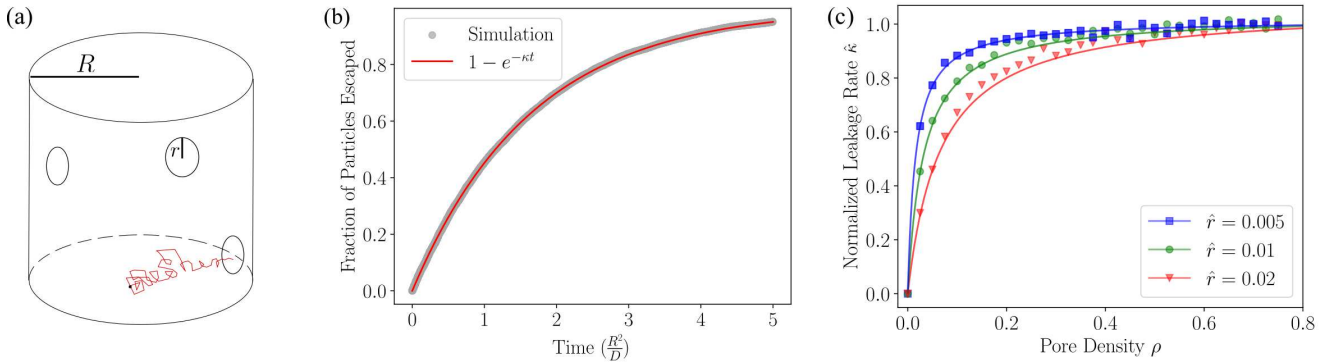}
 	\caption{Simulated escape through small pores in a cylindrical domain. (a) Schematic of the model. The particle is confined in a cylinder of radius $R$ and escapes through pores of radius $r$ on the cylindrical surface.  (b) Effective leakage rate is obtained by fitting a single exponential (red) to the cumulative distribution function of escape times for simulated independent particles (gray).  (c) Normalized leakage rate $\hat{\kappa}=\kappa/\kappa_\text{max}$ relative to the maximum rate of escape, as a function of the pore area density $\rho$. Different colors correspond to different normalized pore sizes $\hat{r}=r/R$. (\emph{Markers}) simulation results; (\emph{lines}) approximate analytical model predictions. }
 	\label{fig1}
 \end{figure*}

 Small absorbing circular pores with radius $r$ are uniformly scattered over the curved surface of the cylinder, with the remainder of the surface reflective (Fig.~\ref{fig1}a). These pores represent the entry sites to channels or pumps that release diffusive particles from the membrane-enclosed tube. For example, they may correspond to Ca$^{2+}$ ion channels on the surface of an ER tube.
 Such channels have a typical pore diameter of approximately $1$nm, giving $r/R \approx 0.01$ for the ER tubule. In the simplest model, we consider the limit where exit is rapid and guaranteed once the particles hit a pore. This limit is relevant for systems with very thin membranes or with active processes that rapidly pull ions through the pore once they have entered. Any such directed forces are assumed to be negligible until the particle has entered the pore itself.
 %
%
This system forms a narrow-escape problem with diffusive particles confined in a cylinder that must exit from scattered small holes. Our goal is to understand the impact of pore density on the rate of escape. 

We implement agent-based simulations for particles moving through this system, employing a hybrid kinetic Monte Carlo (KMC) and Brownian dynamics approach. Analogous approaches have been used to simulate diffusive search of DNA binding proteins \cite{koslover2011theoretical,van2005green}, proteins confined in tubular ER networks \cite{scott2021diffusive}, and reactions of membrane-associated particles~\cite{sokolowski2019egfrd}, as well as general annihilation reactions~\cite{oppelstrup2009first}. In brief, when a particle is far from the wall of the domain, it is propagated to the boundary of the largest possible cylinder centered on its position that does not intersect the domain boundary. The time for such a hop is sampled from the appropriate diffusive propagator function (see Methods). When the particle is near the wall of the cylinder, it is instead moved forward in discrete, small Brownian dynamics time-steps. When such a Brownian step intersects an absorbing pore, the particle is marked as having left the cylinder. This approach allows for efficient propagation of the particles, with large time-steps while they are in the bulk of the domain and small time-steps while they explore near the domain boundary.

 We record the time at which each particle is absorbed and generate an empirical cumulative distribution function, which we then use to find the leakage rate $\kappa$  by fitting to an exponential function $H(t) = 1-e^{-\kappa t}$ (Fig.~\ref{fig1}b). The escape rate is normalized by its maximum value, $\kappa_\text{cyl,max} = 8D/R^2$, which is the inverse of the mean first passage time for uniformly distributed particles to leave a cylinder with a fully absorbing boundary~\cite{redner2001guide}. The simulations are run with different numbers of pores in the cylindrical wall, and the normalized escape rate is plotted as a function of the pore area density ($\rho$, the fraction of surface area that is absorbing).

 As shown previously~\cite{holcman2008diffusion,cheviakov2010asymptotic}, the rate of escape from well-scattered small windows on a domain surface scales linearly with the number of windows. In our simulations, the leakage rate is indeed linear with the area density $\rho$, at small values of $\rho$ (Fig.~\ref{fig1}c). At higher pore area density, the leakage rate approaches the maximum value for a fully absorbing cylinder. When the same area density is split up into larger numbers of smaller pores, the leakage rate is seen to be higher, as has been suggested from asymptotic analysis of general narrow-escape processes~\cite{holcman2015stochastic}.
 
 Notably, the leakage rate rises quite steeply with $\rho$ at low density values. 
  Thus, substantial rates of particle release are expected even with sparsely scattered channels on the cylinder surface. This result mirrors prior estimates for the diffusive current into absorbing patches on an isolated sphere, where sparsely spread small patches were also found to be sufficient for near-maximal absorption~\cite{berg1977physics}. We proceed to develop an analogous approximate model for escape through pores in a cylindrical domain, demonstrating how the leakage rate varies with channel size and density. 
  
 
\subsection{Simplified model predicts escape rate as a function of pore density.}
\label{sec:math}

Approximate calculations for diffusive current into small pores scattered along a surface were first proposed by Berg and Purcell in the context of distant molecules arriving to receptors on the surface of a spherical cell \cite{berg1977physics}. These results highlighted that a small area density of absorbing patches can lead to near-maximal total diffusive current, because particles coming from the bulk explore substantial regions of the surface before wandering back out to infinity. The cylindrical model described here is in many ways analogous, with particles undergoing 3D diffusion until they hit the surface of the confining cylinder, and then exploring along that surface before returning to the bulk. In the regime where $r\ll R$, we would expect the surface exploration to substantially speed up the search process.

\begin{figure*}
	\includegraphics[width=\textwidth]{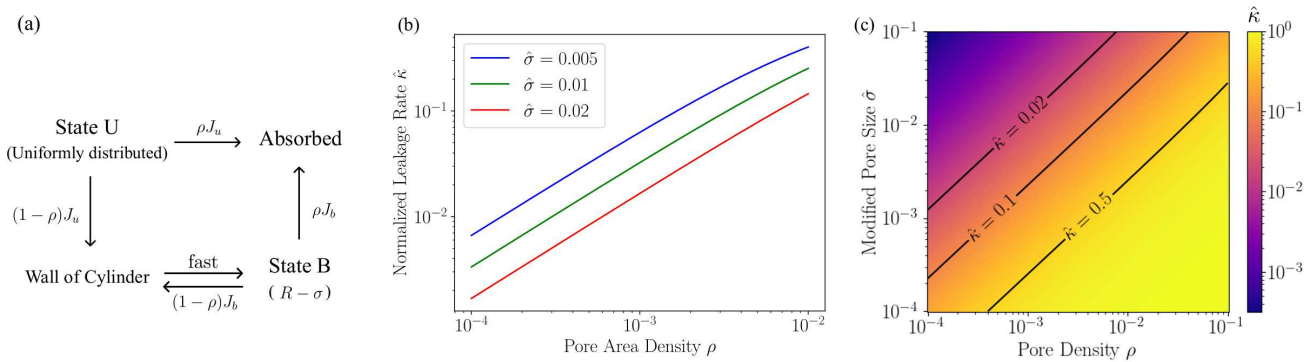}
	\caption{Approximate analytic model for leakage from pores in a cylindrical boundary. (a) State diagram for the simplified model. (b) Normalized leakage rate $\hat{\kappa} = \kappa/\kappa_\text{max}$ at different modified pore sizes $\hat{\sigma}$, for low pore densities $\rho$. (c) Normalized leakage rate $\hat{\kappa}$ as a function of pore area density $\rho$ and modified pore size $\hat{\sigma}$.  Contour curves for several specific leakage rates are shown in black.}
	\label{fig2}
\end{figure*}

We develop an approximate mathematical model for exit through pores in a cylindrical boundary by treating it as a multi-step diffusive process, requiring both finding the surface of the cylinder and searching along it for absorbing pores. The particles can be thought of as transitioning between several distinct states (Fig.~\ref{fig2}a). In state U, particles start uniformly distributed throughout the cylinder and must find the curved cylinder wall. The distribution of transition times for this process is defined as $J_u(t)$. When the particles hit the wall for the first time, the probability that they encounter an absorbing pore is equal to the pore area density $\rho$. The particles that hit a pore immediately leave the domain; reflected particles continue to diffusively explore the cylinder.

 Each sampling of the cylinder surface can be treated as independent, so long as the particle meanders out a preset distance $\sigma$ from the surface of the cylinder, where $\sigma$ should be comparable to the pore size $r$. Berg and Purcell's original model for a patchy absorbing sphere compares the approximate results following this assumption to a complete mathematical solution via an electrostatic analogy, demonstrating that direct correspondence can be achieved by taking $\sigma = (\pi/4)\, r$~\cite{berg1977physics}. We use the same approximation to define $\sigma$ in our model, referring to it as the modified pore size.

  In the limit $\sigma \ll R$, we can neglect the time required to transition from the cylinder surface to the radial position $R-\sigma$. Thus, each reflected particle immediately enters a state B, wherein it starts from radial position $R-\sigma$. The distribution of times to again sample the surface is defined as $J_b(t)$. This transition can include long excursions of the particle into the bulk of the cylinder prior to returning to the boundary. Upon its return to the surface, the particle again has probability $\rho$ of hitting an absorbing pore. 
The particle continues transitioning between the states illustrated in Fig.~\ref{fig2}a until it eventually hits a pore.

The overall distribution of first passage times $J(t)$ for the particle to hit a pore is given by the convolution of transition time distributions over all possible paths that the particle can take between the different states~\cite{wales2002discrete,koslover2011theoretical,koslover2012force}: 
\begin{equation}
    J=\rho J_u+(1-\rho)J_u*\rho J_b+(1-\rho)J_u*(1-\rho)J_b*\rho J_b+...,
\label{eq1}
\end{equation}
where $*$ denotes convolution. The first term corresponds to reaching the pore during the first visit to the surface, the second term includes one reflection event followed by hitting the pore, and so forth. A Laplace transform in the time domain ($t \rightarrow s$) converts the convolutions into products and gives the transformed exit time distribution as:
\begin{equation}
    \hat{J}=\frac{\rho\hat{J}_u}{1-(1-\rho)\hat{J}_b}    
\label{eq:Jcyl}
\end{equation}

The mean time to capture, $\tau$, and the corresponding exit rate ($\kappa = 1/\tau$) can be evaluated from the small-s limit of $\hat{J}(s)$ according to:
\begin{eqnarray}
	\tau & = & -\left.\frac{\partial \hat{J}}{\partial s}\right|_{s=0} \\
	\frac{\kappa}{\kappa_\text{max}} & = & \frac{\rho}{\rho +2\hat{\sigma}(1-\rho)(2-\hat{\sigma})},
	\label{eq:kappacyl}
\end{eqnarray}
where $\hat{\sigma} = \sigma/R$ is the dimensionless modified pore size and $\kappa_\text{max}=8D/R^2$ is the maximal rate for a fully absorbing cylindrical boundary. The normalized leakage rate is shown as a function of the two dimensionless parameters $\rho$ and $\hat{\sigma}$ in Fig.~\ref{fig2}b,c.

The simplified model described here closely reproduces the results of the stochastic simulations (Fig.~\ref{fig1}c), in the narrow-escape regime where $\hat{r}\ll 1$. The approximate analytic calculations begin to underestimate the simulated leakage rate as pores become larger,  due to the incomplete treatment of spatial correlations after each visit to the surface.

 In the limit of low density, the normalized leakage rate is linear with $\rho$, and is given by
\begin{equation}
\begin{split}
\hat{\kappa} & \underset{\rho\rightarrow 0} {\rightarrow} m \rho, \\ 
m & = \frac{1}{2\hat{\sigma}(2-\hat{\sigma})} 
\end{split}
\end{equation}


The slope $m$ indicates how rapidly the leakage rate rises with increasing density of pores. This slope becomes very large for small pore size $\hat{\sigma} \ll 1$, indicating that a given absorbing area fraction can lead to rapid particle release if it is broken up into a large number of very small pores. Conceptually, the diffusing particle samples the surface several times before returning to the bulk interior of the cylinder, and thus has a higher chance of encountering one of a large number of broadly scattered pores.

Considering the leakage of Ca$^{2+}$ from an ER tubule, the approximate pore size is $r \approx 0.5$nm and $\hat{\sigma} \approx 0.008$, corresponding to $m \approx 31$. For this system, reaching a leakage rate that is $1\%$ of maximum would require a pore area density of only $0.03\%$, corresponding to scattering approximately $30$ channels over the surface of a $250$nm-long tubule, or a separation of approximately $\sim 55$nm between neighboring channels (comparable to the size of an ER tubule).
 If we assume calcium diffusivity in the ER lumen to be of the order $D \approx 100\mu\text{m}^2/\text{s}$~\cite{allbritton1992range},
 and the concentration of free luminal calcium to be $\approx 0.5$mM~\cite{yu2000rapid}, this density of channels would give a leakage rate of $k_\text{tot} \approx 2 \times 10^6$ ions per second. Experimental measurements of calcium `blips' in {\em Xenopus} oocytes show that localized (diffraction-limited) regions of ER release ions at an initial rate of roughly $10^6$ ions per second~\cite{sun1998continuum}. Our results imply that a very low area density of calcium channels can be sufficient to account for this rapid release.

\subsection{Domain geometry modulates escape rate dependence on pore density.}
\label{othergeom}

\begin{figure*}
	\includegraphics[width=\textwidth]{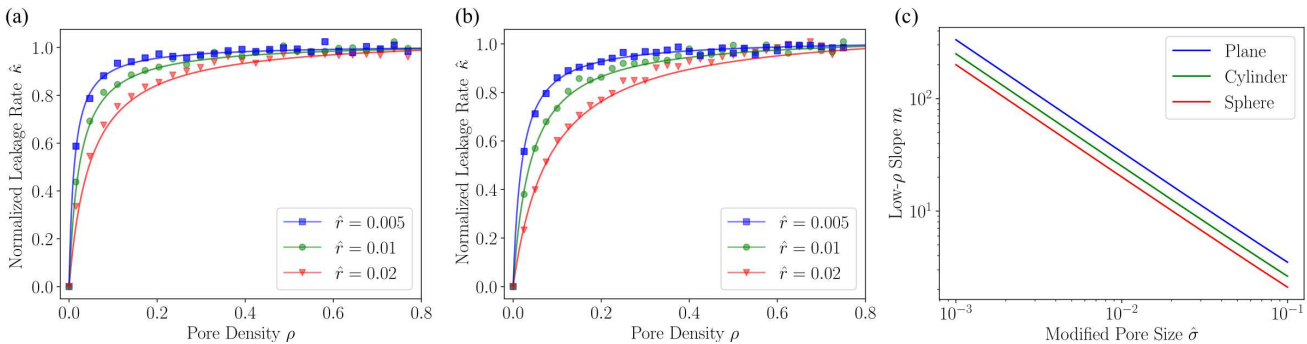}
	\caption{Leakage rate for planar sheets and spherical domains. (a) In planar sheets, normalized leakage rate $\hat{\kappa}$ as a function of the pore area density $\rho$ at different normalized pore sizes $\hat{r}$. (\emph{Markers}) simulation results; (\emph{lines}) analytic model predictions. (b) Corresponding results for a spherical domain. (c) Slope ($m$) of leakage rate with increasing pore density, in the low $\rho$ limit, plotted as a function of modified pore size $\hat{\sigma}$ in different geometries. }
	\label{fig3}
\end{figure*}

The simulation approach and analytical model described above can be extended to a variety of different geometries for the membrane-enclosed domain. For example, we consider particles escaping through holes in a planar sheet  between parallel flat membranes (as found in the perinuclear ER~\cite{schwarz2016endoplasmic}) or in a membrane-enclosed sphere (including globular organelles such as peroxisomes~\cite{antonenkov2012transfer} and lysosomes~ \cite{li2019lysosomal,yang2019release,patel2015evolution}). 


The primary difference between these systems is the dimensionality within which the confined particle diffuses to encounter the membrane: 1D for sheets, 2D for cylinders, and 3D for spheres. In each case, upon reflecting from the wall, the particle will follow diffusive paths that explore the domain further before resampling the surface. 


Using the same approximate analytic approach as described for cylinders, the general normalized leakage rate for these models can be expressed as
\begin{equation}
    \frac{\kappa_\nu}{\kappa_{\nu, \text{max}}} = \frac{2\rho}{2\rho+(2+\nu)\hat{\sigma}(1-\rho)(2-\hat{\sigma})}
    \label{eq:kappagen}
\end{equation}
where $\nu=1,2,3$ for sheets, cylinders, and spheres, and the maximum leakage rate is $\kappa_{\nu,\text{max}} = 3D/R^2$, $8D/R^2$, and $15D/R^2$, respectively.

In Fig.~\ref{fig3}a,b we show that the analytical model approximately matches simulations for diffusing particles in both sheets and spheres with scattered small pores. As in the cylindrical case, the  rate is linearly dependent on pore area density at low values of $\rho$, with slope

\begin{equation}
    m_\nu = \frac{2}{(2+\nu)\hat{\sigma}(2 -\hat{\sigma})}
\end{equation}

 The slope is higher in planar sheets and lower in spheres (Fig.~\ref{fig3}c).  
  Thus, in a flat sheet, sparser pores are sufficient to reach a comparable fraction of the maximum leakage rate, as compared to a sphere. This effect can be explained by noting that the average time to wander away from a confining spherical boundary towards the bulk of the sphere is longer than the time to wonder away from a planar sheet, due to the effective outward radial bias associated with diffusion in three dimensions~\cite{redner2001guide}. Thus, in the spherical case, the particle is able to more thoroughly explore the surface with each visit, rendering the exit time less steeply dependent on the pore density.
  


\subsection{Channel length lowers escape rates and pore density dependence.}
\begin{figure}
	\includegraphics[width=8cm]{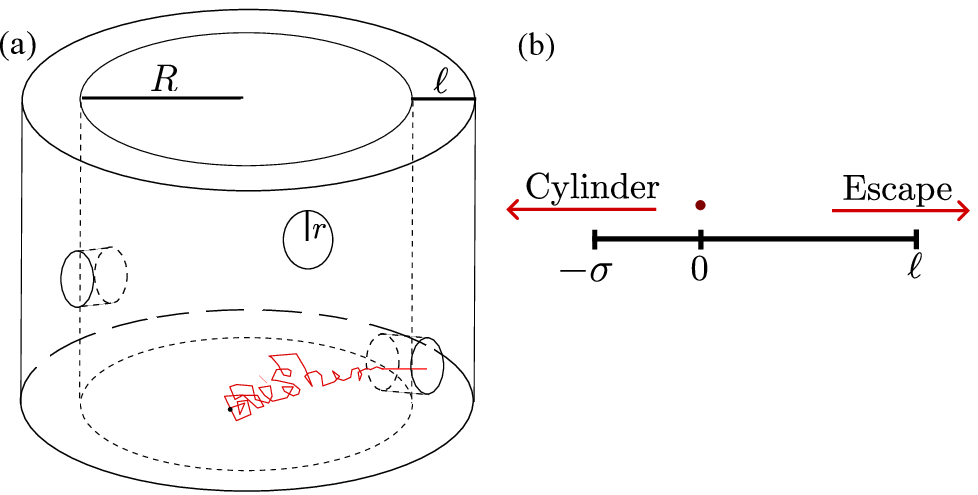}
	\caption{Schematic for model with non-zero channel length. (a) In the bulk cylidnrical domain, the particle diffuses to encounter the entrance of a channel on the domain boundary with radius $r$, followed by entrance and diffusion through the channel of length $\ell$ to escape. (b) Within the channel, the particle undergoes effectively 1-dimensional motion, either to escape the domain or return to the cylinder. }
	\label{fig4}
\end{figure}

We next proceed to consider the case where the particle has to make its way through a channel of non-zero length after entering the pore. We assume the motion of the particle in the narrow channel is diffusive and one-dimensional, that the channel has length $\ell$, and that the particle can leave the channel either by re-entering the confining domain or by escaping the system entirely through the far end of the channel (Fig.~\ref{fig4}).

\begin{figure*}
	\includegraphics[width=\textwidth]{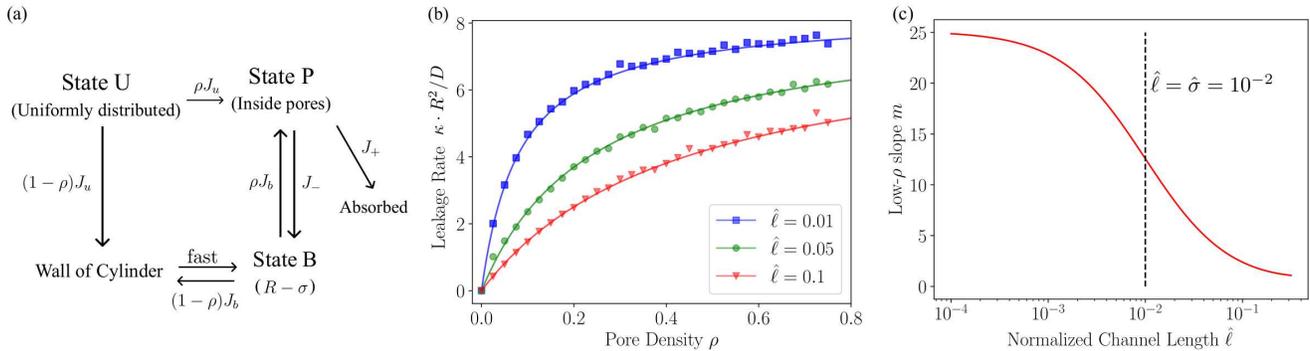}
	\caption{Model for escape from a cylinder through channels of non-zero length. (a) State diagram for the simplified model incorporating channel length. (b) Leakage rate $\kappa$  as a function of the pore area density $\rho$ at different channel lengths $\hat{\ell}$. (\emph{Markers}) simulation results; (\emph{lines}) simplified model predictions. Rate is not normalized to highlight slower escape with longer $\ell$. (c)  Slope $m$ relating normalized escape rate to channel density $\rho$ in the low-$\rho$ limit, plotted as a function of normalized channel length $\hat{\ell}=\ell/R$. The slope decreases substantially when the channel length becomes comparable to the modified pore size $\hat{\ell}\approx \hat{\sigma} = 10^{-2}$ (\emph{black line}).}
	\label{fig5}
\end{figure*}

 We propose a similar method to our previous analytical model to approximately describe the leakage rate in a system with non-zero channel length.  For simplicity, we focus specifically on cylindrical domains, although analogous calculations can also be made for planar sheets and spheres. An extra state (state P) describes a particle that has just entered the pore, either from a uniform disribution (state U) or from a starting point at distance $\sigma$ from the domain wall (state B), as schematized in Fig.~\ref{fig5}a. The distributions of transition times into state P remain the same as in the previous model.  Once in state P, the particles perform 1D diffusion in the pore, starting at $x=0$ between absorbing boundaries at position $x_1=-\sigma$ and $x_2=\ell$. Absorption at the first boundary (transition time distribution $J_-$) corresponds to leaving the pore and entering back into the cylinder, moving a sufficient distance ($\sigma$) away from the boundary to make an effectively independent attempt at sampling the domain surface. Absorption at $x_2$ (time distribution $J_+$) corresponds to successfully escaping the domain.
 As in the previous model, the convolutions of transitions over all possible paths are summed to give the overall escape time distribution. The Laplace-transformed escape time distribution is then: 
 
\begin{equation}
\begin{split}
	\hat{J}(t)& =\frac{\rho\hat{J}_u\hat{J}_+}{1-(1-\rho)\hat{J_b}-\rho\hat{J_b}\hat{J}_-}.
    \label{eq:thick1}
\end{split}
\end{equation}
Plugging in the expressions for individual time distributions, for cylindrical confinement, we have:

\begin{equation}
\begin{split}
    \frac{\kappa}{\kappa_\text{max}}=\frac{\rho\(2\hat{\sigma}\hat{\ell}+(2\hat{\ell}+1)^2\)}
    {\rho+2\hat{\sigma}(1-\rho)(2-\hat{\sigma})+2\hat{\ell}\(2\rho\hat{\sigma}+2\rho\hat{\ell}+2-\hat{\sigma}\)}
    \label{eq:thick2}
\end{split}
\end{equation}
where $\kappa_\text{max}=8D/(2\sigma \ell+(2\ell+R)^2)$ is the maximal escape rate for a surface fully covered with channels. It should be noted that this model is only accurate when $0<\ell\ll R$, which is suitable for most biological applications. For instance, the length of the pore in ion channels is comparable to the thickness of the membrane ($\sim 4$nm)\cite{prasad2020mapping}, and much smaller than the radius of an ER tubule ($R \approx 50$nm). 
In the limit $\ell\rightarrow 0$, the exit rate approaches Eq.~\ref{eq:kappacyl}, as expected. 

The approximate model described here shows good agreement with agent-based stochastic simulations (Fig.~\ref{fig5}b). Similar to previous sections, we are interested in the dependence on channel density $\rho$ as described by the slope $m$ in the low-$\rho$ limit. For the case with channels of non-zero length, this slope is given by,
\begin{equation}
    m=\frac{2\hat{\sigma}\hat{\ell}+(2\hat{\ell}+1)^2}{2(2-\hat{\sigma})(\hat{\ell}+\hat{\sigma})}.
\end{equation}

The introduction of channel length does not substantially affect the system when $\hat{\ell}$ is extremely small, as particles that enter a pore are likely to quickly escape. However, as $\hat{\ell}$ increases to be comparable to $\hat{\sigma}$, the slope $m$ becomes dramatically smaller (Fig.~\ref{fig5}c). For the example system of Ca$^{2+}$ ions diffusing out of pores in an ER tubule, we can estimate $\hat{\sigma} \approx 0.008$ and $\hat{\ell} \approx 0.08$. The predicted slope of the normalized exit rate versus pore density is reduced to $m\approx 3.8$, as compared to the steep dependence on pore density ($m\approx 31$) in the case of instantaneous exit through pores. In order for a $250$nm segment of tubule containing $0.5$mM Ca$^{2+}$ to release ions at a rate of $10^6 \text{s}^{-1}$, this estimate would imply a channel area density of $\rho \approx 0.2\%$, corresponding to a separation of $23$nm between neighboring channels, or to a total of approximately $180$ channels in the tubule section. 

The overall escape rate can also be converted to an effective permeability of the domain boundary, according to $P = R\kappa/2 \rightarrow m\rho R\kappa_\text{max}/2$. For the example system of calcium escape from an ER tubule, our estimates imply an effective permeability of $P \approx (2.3\text{cm/s}) \rho$ or $P \approx 40\mu\text{m/s}$.

It should also be noted that this calculation assumes that ions entering the pore must diffuse passively within it and that they are capable of escaping back to the ER lumen rather than going all the way through the pore. In the case of active transport mechanisms or potential gradients within the channel structure, a more accurate model may well be the limiting case covered in the previous section where entrance into the pore guarantees rapid exit out of the tubule.

\section{Conclusions}
\label{sec:conclusion}

This work presents simple expressions for the diffusive leakage rate of particles escaping through pores or channels in membrane-enclosed domains. We use an approximate approach that breaks the escape process into steps between effective discrete states wherein a particle is in the bulk of the domain, exploring near the membrane, or passing through the pore. Analogous approaches have been previously employed to model diffusive search in a number of different systems, including chemoreception on the surface of a sphere~\cite{berg1977physics}, facilitated diffusion of transcription factors searching for DNA binding cites~\cite{koslover2011theoretical}, and proteins encountering exit sites in the peripheral ER network~\cite{scott2021diffusive}. We use propagator-accelerated agent-based simulations~\cite{oppelstrup2009first,koslover2011theoretical, scott2021diffusive} to validate our analytic approximations. Good agreement is found between the mathematical results and simulations for cylinder, sphere, and planar sheet domains. 

These results delineate the dependence of the effective current out of the domain on the pore size, the density of pores, and the length of channel which must be traversed upon entering the pore. For the case where escape is rapid and guaranteed upon pore entry, we find that leakage rate rises steeply with the pore density $\rho$, approaching high fractions of the maximum possible escape rate even when the absorbing pores cover a small fraction of the domain boundary. 
 For such systems, adding a few well-scattered pores over the membrane surrounding a domain can greatly enhance the effective permeability of that membrane.
 This dependence on density is particularly steep for small pore sizes ($\hat{\sigma}\rightarrow 0$), implying that breaking up the same absorbing area into large numbers of smaller pores will lead to faster diffusive escape.


Our simplified model is extended to consider the effect of the channel length $\ell$ that must be traversed after entering a pore. For passive channels, this length should be at least as long as the membrane thickness. For low channel densities, we show that the overall leakage rate becomes much slower when the channel length is comparable to or greater than the pore size. As the channels become longer, the pores must cover an increasingly large area fraction of the boundary in order to yield substantial flux out of the domain. Notably, our model assumes that transport through the channel upon entering the pore consists of unbiased one-dimensional diffusion. The effective rate of channel transit could be substantially sped up by potential gradients that drive ions through the channel or by active processes facilitating the transit.

Overall, these results provide an approach for estimating the effective membrane permeability in a region with a certain density of pores or channels. We consider as our example case the release of calcium ions from tubules of the peripheral endoplasmic reticulum. In certain cell types, the ER locally releases a bolus of calcium ions into the cytoplasm, referred to as puffs, blips, twinkles, or sparks~\cite{sun1998continuum,cheng1993calcium,kanemaru2014vivo}. Such release requires a small (sub-micron) localized region of the ER membrane to open calcium channels~\cite{sun1998continuum,smith2009imaging,lohmann2005local} and become permeable to the Ca$^{2+}$ ions stored in the ER lumen. The interplay of timescales between luminal calcium transport and local release~\cite{konno2021endoplasmic} depends on the overall permeability of the releasing region. Although the density of channels that open during a release event is unknown, our estimates indicate that only a few dozen channels scattered over a short region of tubule can be sufficient to yield the observed rate of calcium efflux~\cite{sun1998continuum}. However, if the calcium release is limited by diffusive transport through the channel itself then approximately a hundred channels would be needed within the same tubule to give the same overall flux.
 Future experimental measurements aiming to quantify the density of calcium channels near local release regions would be required to test the estimations provided here.

Further extension of this model could include consideration of more complicated domain geometries. The approach of reducing diffusive search times to transit between effective discrete states has been previously used successfully for modeling complex architectures such as folded genomic structures~\cite{koslover2011theoretical} and tubular networks~\cite{scott2021diffusive}. Other cellular systems where diffusive particle escape can play an important role include mitochondria, with the rows of involuted cristae comprising their inner membrane~\cite{cogliati2016mitochondrial} and the outer segment of photoreceptor cells with their stacks of pancake-like membranous discs~\cite{nickell2007three}. The escape of proteins from peripheral ER sheet regions into adjoining tubules could also be formulated as an analogous problem  for diffusive escape from a flat disc. Of further interest
is the possibility of non-uniform diffusion for particles passing near the pores. Specifically, many cellular structures (such as nuclear pores~\cite{wente2010nuclear} and possibly ER exit sites~\cite{barlowe2016cargo}) are gated by regions with some amount of binding affinity to the passing particles. These regions can form aggregates or liquid droplet-like structures that may enhance the ability of diffusive particles to enter the pores~\cite{heltberg2021physical,celetti2020liquid}. Exploring how a multi-faceted search process involving finding first these gating structures and then the pore entry itself could prove a fruitful avenue for future work.

Overall, the calculations presented in this paper highlight the non-trivial relationship between channel density, membrane thickness, and effective membrane permeability. The high permeability values that arise even for relatively low channel densities make it physically possible for a given membrane to be packed by many different channels responsible for the selective release of different particles, while still maintaining efficient release rates. Theoretical exploration of the interplay between channel parameters and escape rates yields important insights into a ubiquitous process in cell biology -- the release of particles from membranous intracellular structures.
\vfill



\section{Methods}

\subsection{Simulations}

\subsubsection{Simulating escape times from pores in a cylinder}
In each of our agent-based simulations,  point-like non-interacting particles attempt to escape a cylindrical domain through small pores scattered on its surface. The particles were initiated with a uniform distribution throughout the cylinder.
The top and bottom of the cylinder were treated as periodic boundary conditions.
 The height of the cylinder did not affect the result.

 We employed a hybrid kinetic Monte Carlo (KMC) and Brownian dynamics approach. Whenever a particle was sufficiently far from the domain boundary, it was propagated through time and space by 
 %
   considering the widest possible cylinder centered at its position yet completely contained within the domain. We sampled the time to first encounter the radial boundary of this internal cylinder from the known diffusive Green's function in cylindrical coordinates~\cite{carslaw1947conduction}. Specifically, the distribution of first passage times to an absorbing cylinder of radius $b$ is given by
\begin{equation}
\begin{split}
    \psi_\text{cyl}(t; b)=\frac{1}{\pi b^2}\sum_{n=0}^{\infty} \frac{J_0(r\frac{\alpha_n}{b})}{J_1(\alpha_n)^2}\exp(-D\frac{\alpha_n^2}{b^2}t),
\end{split}
\end{equation}
where $J_0$ and $J_1$ are the zeroth and first order Bessel function of the first kind and $\alpha_n$ is the n-th root of $J_0$. 
We used the inverse cumulant method to sample from this distribution of times after finding the cumulative survival probability $S(t)$:
\begin{equation}
\begin{split}
    S_\text{cyl}(t)=\int_{0}^{R} 2\pi R\psi(r,t)dr=2\sum_{n=0}^\infty\frac{\exp(-D\frac{\alpha_n^2}{b^2}t)}{\alpha_n J_1(\alpha_n)}.
\end{split}
\end{equation}
After each such propagation step, the time counter for the particle was updated by the sampled value $t_\text{samp}$. The new angular position of the particle on the surface of the propagating cylinder was selected uniformly, and the axial position was chosen from a normal distribution $\mathcal{N}(0,\sqrt{2 D t_\text{samp}})$, with periodic boundary conditions.

This approach of propagating particles within `protected domains' using an analytically solved propagator has been referred to as `Green's function reaction dynamics'~\cite{van2005green,sokolowski2019egfrd} or `first passage kinetic Monte Carlo'~\cite{oppelstrup2009first}, and has been employed to simulate a variety of systems where it is efficient to match the time-step to the extent of the empty space around the particle~\cite{koslover2011theoretical,sokolowski2019egfrd,scott2021diffusive,redner2001guide}.

When the particle approached close to the domain boundary, it propagated in 3D Brownian steps, sampled from a normal distribution with standard deviation  $\sqrt{2D\Delta t}$ in each dimension. The time-step was taken to be $\Delta t \approx 10^{-8} R^2/D$ (time-steps one order of magnitude smaller did not noticeably affect the results). The distance cutoff for switching to Brownian steps (versus KMC propagation) was $\sqrt{2D\Delta t}$.

Whenever a Brownian step took the particle outside the domain boundary, the crossing-point on the boundary was computed. If this crossing point was within a pore on the domain boundary, the particle was considered to have left the cylinder. Otherwise, the new position of the particle was radially reflected back into the domain.

The simulations were run for $10^4$ independent particles, up to a maximum time of $t = 3 R^2/D$, or until all particles have left the domain. A variety of pore densities and sizes were considered. Each pore was described as a circular patch of radius $r$, curved to conform to the surface of the domain cylinder. The pore centers were placed at equispaced lattice locations around the domain boundary, allowing for a wide range of densities. 

To get the leakage rate $\kappa$ from each run, we fit the cumulative distribution of escaped particles to an exponential function $H(t)=1-e^{-\kappa t}$, or, if all particles had escaped, we calculated the mean time to capture and take its inverse. 

\subsubsection{Escape times from plane sheet and sphere}

We employed similar methods in simulations where the domain consisted of a plane sheet (enclosed by 2 flat rectangular boundaries with periodic boundary conditions) or of a spheres, both with circular pores scattered on their surface. 

For the plane sheets, the z-axis was defined normal to the two parallel domain boundaries, and the distance between the boundaries was set to $2R$.
 The length and width of the sheet does not affect the system as long as the dimensions are much larger than the pore size. 
 
 As in the cylindrical case, whenever the particle was far from the rectangular domain boundaries, it was enclosed in a protective slab of height $b$ centered on its $z$ position and extending as far as the nearest domain boundary. A kinetic Monte Carlo step was used to sample the time for the particle to leave this slab, from the known survival probability of a particle undergoing 1D diffusion between two absorbing boundaries: 
 %
\begin{equation}
\begin{split}
    S_\text{plane}(t)=\frac{4}{\pi}\sum_{n=1}^\infty \frac{(-1)^{n+1}}{2n-1}\exp\(-D\frac{(2n-1)^2\pi^2}{b^2} t\)
\end{split}
\label{eq:Splane}
\end{equation}

The $x$ and $y$ position of the particle were selected from a normal distribution based on the sampled time to leave the slab $\mathcal{N}(0,\sqrt{2Dt_\text{samp}}))$. As before, discrete Brownian dynamics steps were used to propagate the particle whenever its position was within $\sqrt{2D\Delta t}$ of the domain boundary.

In the spherical model, a particle far from the boundary was propagated to the largest possible sphere (radius $b$) centered at the particle's position and contained within the spherical domain. The survival probability for the spherical propagation is given by
\begin{equation}
\begin{split}
    S_\text{sphere}(t)=2\sum_{n=1}^\infty (-1)^{n+1}\exp(-D\frac{n^2\pi^2}{b^2} t)
\end{split}
\end{equation}
After each KMC step, the particle was placed at a random angular position on the surface of the propagation sphere.

\subsubsection{Channel length}
In the cylindrical model with channel length, a particle that enters a pore has to diffuse through a channel of length $\ell$ to successfully escape from the domain. We modeled diffusion inside a channel to be effectively 1-dimensional, as the channel radius was assumed much smaller than the domain.

Propagation of the particle within the channel was done through a hybrid KMC and Brownian dynamics method. Whenever the particle was far from the two channel boundaries, its step was sampled from the survival probability for 1D diffusion between absorbing bounds (Eq.~\ref{eq:Splane}). Whenever it approached within $\sqrt{2D\Delta t}$ of either boundary, it was propagated in discrete Brownian steps. The time-step used was the same as in simulations without channel length. When the particle left the channel through the outer boundary, it was marked as having escaped. When it left the channel through the inner boundary, it was placed back in the cylindrical domain, with its radial position determined by the end-point of the Brownian step bringing it out of the pore, and its angular and axial positions selected uniformly within the pore's cross-section. The particle then proceeded to explore the cylinder until another Brownian step caused it to enter a pore.


\subsection{Simplified analytic model}

\subsubsection{Arrival to pores}

For the cylindrical model, we can directly compute the Laplace-transformed transition time distributions between the states illustrated in Fig.~\ref{fig2}a. Specifically, the first passage time distribution for hitting the domain boundary starting from a uniform distribution ($J_u$) and  from radial position $R-\sigma$ ($J_b$) can be found by solving the diffusion equation in cylindrical coordinates~\cite{carslaw1947conduction, redner2001guide}. They are given, respectively, by:
%
%
\begin{eqnarray}
    \hat{J_u} & = & \frac{2\sqrt{\frac{D}{s}}}{R}\frac{I_1(R\sqrt{\frac{s}{D}})}{I_0(R\sqrt{\frac{s}{D}})} \\
    \hat{J_b} & = &\frac{I_0((R-\sigma)\sqrt{\frac{s}{D}})}{I_0(R\sqrt{\frac{s}{D}})},
\end{eqnarray}
where $I_0$ and $I_1$ are zeroth and first order modified Bessel functions of the first kind. These transition time distributions can be convolved together to obtain the overall first passage time distribution for exiting the domain ($\hat{J}$), as given in Eq.~\ref{eq1}-\ref{eq:Jcyl}.

Computing the mean time to capture ($\tau=-\left.\frac{\partial \hat{J}}{\partial s}\right|_{s=0}$) requires only the first two terms in the Taylor expansions of $\hat{J}_u$ and $\hat{J}_b$ with respect to $s$, which are,
\begin{equation}
\begin{split}
    \hat{J_u} & = 1-\frac{R^2}{8D}s+O(s^2) \\
    \hat{J_b} & = 1-\frac{2\sigma R-\sigma^2}{4D}s+O(s^2)
\end{split}
\end{equation}

After plugging in the expressions and taking the partial derivative with respect to $s$, we find,
\begin{equation}
\begin{split}
    \tau & =\frac{R^2}{8D}+\frac{(1-\rho)(2\sigma R-\sigma^2)}{4D\rho}
\end{split}
\end{equation}
By taking the inverse, we obtain Eq.~\ref{eq:kappacyl}. 

The Laplace-transformed exit time distributions in other geometries can be found in an analogous way. 
For planar sheets, we have the transition time distributions:
\begin{eqnarray}
\hat{J}_{u,\text{plane}} & = & \frac{1}{R\sqrt{sD}}\frac{\exp(2R\sqrt{\frac{s}{D}})-1}{\exp(2R\sqrt{\frac{s}{D}})+1} \\
\hat{J}_{b,\text{plane}} & = & \frac{\sinh((2R-\sigma)\sqrt{\frac{s}{D}})+\sinh(\sigma\sqrt{\frac{s}{D}})}{\sinh(2R\sqrt{\frac{s}{D}})}.
\end{eqnarray}
The corresponding mean first passage time is
\begin{eqnarray}
\tau_\text{plane} & = & \frac{R^2}{3D}+\frac{(1-\rho)(2\sigma R-\sigma^2)}{2D\rho}.
\end{eqnarray}

For a spherical domain, the transition time distributions are:
\begin{eqnarray}
\hat{J}_{u,\text{sphere}} & = & \frac{3D\(R\sqrt{\frac{s}{D}}\cosh(R\sqrt{\frac{s}{D}}\)-\sinh\(R\sqrt{\frac{s}{D}})\)}{sR^2\sinh(R\sqrt{\frac{s}{D}})} \\
\hat{J}_{b,\text{sphere}} & = & \frac{R\sinh((R-\sigma)\sqrt{\frac{s}{D}})}{(R-\sigma)\sinh(R\sqrt{\frac{s}{D}})},
\end{eqnarray}
with the mean escape time
\begin{eqnarray}
\tau_\text{sphere} & = & \frac{R^2}{15D}+\frac{(1-\rho)(2\sigma R-\sigma^2)}{6D\rho}.
\end{eqnarray}
The inverse of these escape times yields the leakage rate expression in Eq.~\ref{eq:kappagen}.


\subsubsection{Escape from channels of non-zero length}
In systems where the particle must exit the pore through a channel of length $\ell$, two new transition time distributions are required. Specifically, $J_+$ and $J_-$ represent the transition time distribution from state P (having just entered the pore) to exit on the far side and to state $B$ (return to the domain at radial position $R-\sigma$), respectively. The diffusive process inside a channel is modeled as a 1D diffusion starting at $x=0$, with two absorbing boundaries at $x=-\sigma$ and $x=\ell$. The flux out of this 1D region through each boundary is given by~\cite{redner2001guide}:
\begin{equation}
\begin{split}
    \hat{J}_+ & =\csch\(\sqrt{\frac{s}{D}}(\ell+\sigma)\)\sinh\(\sqrt{\frac{s}{D}}\sigma\) \\
    \hat{J}_- & =\csch\(\sqrt{\frac{s}{D}}(\ell+\sigma)\)\sinh\(\sqrt{\frac{s}{D}}\ell\)
\end{split}
\end{equation}
Taking the Taylor expansion, and keeping terms up to linear order gives:
\begin{equation}
\begin{split}
    \hat{J}_+ & = \frac{\sigma}{\ell+\sigma}-\frac{2\ell\sigma^2+\ell^2\sigma}{6D(\ell+\sigma)}s+O(s^2) \\
    \hat{J}_- & = \frac{\ell}{\ell+\sigma}-\frac{\ell\sigma^2+2\ell^2\sigma}{6D(\ell+\sigma)}s+O(s^2)
\end{split}
\end{equation}


The overall rate of leaving the domain out of the channel involves a convolution of transitions back and forth between different states schematized in Fig.~\ref{fig5}a. The sum of the Laplace-transformed convolutions over all paths can be expressed as
\begin{equation*}
\begin{split}
\hat{J} & = \left[ \rho \hat{J}_u \hat{J}_- + (1-\rho) \hat{J}_u\right] \left[\sum_{n=0}^\infty \left((1-\rho) \hat{J}_b + \rho \hat{J}_b\hat{J}_-  \right)^n\right] \times \\
& \times \rho \hat{J}_b \hat{J}_+ + \rho \hat{J}_u \hat{J}_+.
\end{split}
\end{equation*}
The first term corresponds to the particle either hitting a pore (state P) and then returning to the bulk of the cylinder (state B) or hitting the reflecting wall. The summation term incorporates all path loops starting and ending at state B, where $n$ is the total number of loops and each loop can involve either an excursion to the reflecting wall or to a pore and back again. Each path that passes through state B must eventually return to a pore and exit. Finally, the last additive term corresponds to paths where the particle enters the pore and exits directly without ever visiting state B. Simplifying this expression gives the overall Laplace transformed distribution for exiting through a channel (Eq.~\ref{eq:thick1}). By plugging in the previous expressions of individual transition time distributions, we arrive at Eq.~\ref{eq:thick2}.

\bibliography{channelLeakRate} 

\end{document}